# On Properties of Phase-Conjugation Focusing for Large Intelligent Surface Applications – Part I: Vertical Polarization

Jiawang Li[1]

[1] Department of Electrical and Information Technology, Lund University, 221 00 Lund, Sweden

jiawang.li@eit.lth.se

*Abstract*—Large intelligent surface (LIS) is one promising path to leverage 6G performance in sub-10 GHz bands. This two-part paper explores the properties of phase-conjugation focusing for a simplified LIS setup with a two-dimensional (2D) circular antenna array and a user antenna located within the array aperture in the same plane. In Part I of this article, we assume vertical polarization for all antenna elements, whereas Part II assumes horizontal polarization. In Part I, we focus on the effect of array radius on the peak gain, 3 dB focusing width, and sidelobes for two types of circular arrays. The numerical results show that the gain minimum is located at the array center. The peak gain varies by less than 0.5 dB for focal points located within $2\lambda$ from the array center. Similarly, the focal width and sidelobe level are also stable within this region, irrespective of array radius. From $2\lambda$ from the center to the array edge, the closer proximity of the focal points to some array elements than other elements results in more drastic changes in these NFF properties. Finally, full-wave simulation using Ansys HFSS is used to partially validate the numerical results.

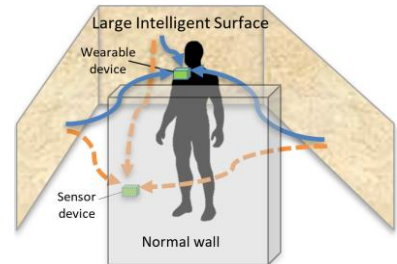

Fig. 1. Examples of LIS applications: wireless communications and wireless power transfer.

## I. Introduction

Large intelligent surface (LIS) [1] is a promising concept for fulfilling tough 6G requirements in sub-10GHz bands, given the limited frequency spectrum resource. A LIS consists of an electrically large active surface that transmits and receives electromagnetic waves. As depicted in Fig. 1, when the walls of a room are covered with LISs, they can facilitate not only communication but also other applications including power transmission and positioning to wearable devices on the human body as well as various indoor sensor devices. Since LIS is practically implemented with a large number of antenna elements widely distributed in space, it can be seen as a form of distributed MIMO, unlike the 5G massive MIMO concept with a large number of localized antenna elements.

Given the wide distribution of array elements, LIS goes beyond the concept of far-field (FF) beamforming (e.g., in 5G) to also include scenarios of near-field-focusing (NFF), where the user may be in the radiative near-field (NF) region of the array elements. The working principles and applications of NFF, including the methods for shaping the focal region have been investigated [2]-[5]. NFF methods can be classified into: 1) phase conjugation method [3] (similar to classical FF delay-and-sum beamformer) and 2) multi-objective optimization methods [4]. The phase conjugation method has a clear physical meaning (i.e., delay-and-sum) and its array weights can be obtained in closed form for the simple case of no scattering (i.e., line-of-sight (LOS) propagation channel) or using estimated channels in real scenarios. However, it only offers one set of array weights with some corresponding focusing properties. On the other hand, optimization methods facilitate the flexible shaping of focal region(s) by adjusting the signal phase and amplitude using array weights, often still relying on LOS model. The drawbacks are the lack of physical insight, potentially slow convergence, and distortion from scattering in real channels.

Due to the simplicity of the phase conjugation method, its NFF properties have been explored in traditional NFF cases where a single large planar array is used and the focal point being in the array's radiative NF region. A previous study provided a representative list of NFF properties [3]:

1) Focal gain: The focal gain increases with the array's electrical size ($L/\lambda$), where $L$ is the largest array dimension and $\lambda$ the free space wavelength. For large arrays ($L/\lambda > 20$), it approaches the theoretical limit of continuous-aperture antennas.
2) Focal size: The 3 dB focal diameter in the focal plane (parallel to array plane) decreases with increasing electrical size of the array and is approximated as $\Delta_s \approx 0.866\, r_0/L$, where $r_0$ is the distance from the array to the focal plane. This result can be compared with the 3 dB FF beamwidth being proportional to array aperture. The depth of focus (DoF), defined as the axial range where power density drops to 3 dB below its peak, increases with increasing focal distance.
3) Focal shift: The peak power density in the NF does not occur exactly at the designed focal point but shifts toward the array.

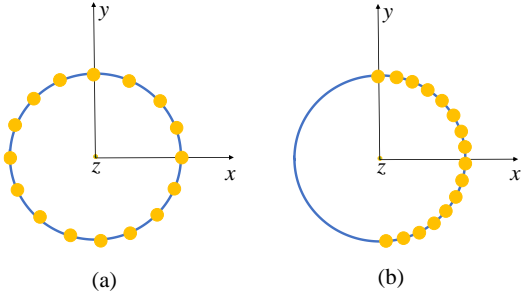

Fig. 2. Two types of UCAs: (a) full circle, (b) half-circle. There are 16 array elements for each type in this example.

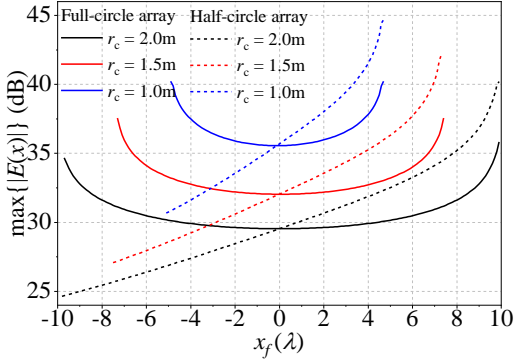

Fig. 3. Peak gain of full-circle and half-circle UCAs.

4) Spurious focal regions: Array element spacing of above $1\lambda$ introduces spurious focal region(s), which is akin to grating lobes in FF patterns with array spacing over $\lambda/2$.

As explained, whereas some of these NFF properties find close parallels to FF beamforming, others are unique to traditional NFF scenarios [2]-[5].

In this two-part paper, we consider the more general NFF scenario where the user is surrounded by LIS elements from more than one side (e.g., see Fig. 1). Specifically, circular arrays are used as the reference LIS array configuration and the user is "inside" the LIS aperture. Furthermore, the effect of polarization is explicitly included, with Part I of this article dealing with vertical polarization and Part II with horizontal polarization.

## II. NFF CHARACTERISTICS OF CIRCULAR ARRAYS

To keep the study simple, we consider the LIS setup with uniform circular array (UCA) and phase conjugation NFF. We apply several simplifying assumptions of the array elements: negligible mutual coupling, omnidirectional pattern in the $xy$ plane, and vertically polarized (along the $z$-axis). The user's antenna is also omnidirectional in the $xy$-plane and vertically polarized. These assumptions can be fair approximations of real cases, e.g., vertical dipole antennas with element spacing of $\lambda/2$ or larger. Here, it is further assumed that there is no scattering object nor any scattering by the array elements. The effect of antenna scattering will be evaluated in Section III.

We consider two types of $N$-element UCAs with radius $r_c$, i.e., full vs. half circle (see Fig. 2). The element spacing of the second type is roughly half of the first type. This can be a relevant comparison, e.g., if antenna selection is used.

With phase conjugation, the phasor form of the ($z$-polarized) electric FF at the observation point $\boldsymbol{r} = (x, y)$ is given by [4]

$$\boldsymbol{E}(\boldsymbol{r}) = \boldsymbol{a}_z E_0 \sum_{n=1}^{N} \frac{e^{-j2\pi(|\boldsymbol{r}-\boldsymbol{r}_n|-|\boldsymbol{r}_f-\boldsymbol{r}_n|)/\lambda}}{|\boldsymbol{r}-\boldsymbol{r}_n|} \text{ V/m}, \quad (1)$$

where $\boldsymbol{r}_n$ is the position vector of element $n$, $\boldsymbol{r}_f$ is the position vector of the focal point, and $E_0$ is the initial field amplitude at each element. We assume $E_0 = 1$ V/m. If $N$ is even, $x_n = r_c \cos\frac{2\pi(n+1)}{N}$ and $y_n = r_c \sin\frac{2\pi(n+1)}{N}$. If $N$ is odd, $x_n = r_c \cos\frac{2\pi n}{N}$ and $y_n = r_c \sin\frac{2\pi n}{N}$. It is noted that when applying (1) to the radiative NF region of this setup, the waves from different array elements arrive at the receiving dipole from different directions, unlike the FF case. However, the presence of the antenna (with omnidirectional pattern in the $xy$ plane) allows the wave propagation directions to be neglected, unlike superposition of the waves' electric fields in free space, where cancellation can occur simply due to opposing wave directions. Therefore, (1) is valid for the received electric field at the antenna even for the NF region.

In this study, we focus on three circular arrays of each type with $r_c = 1$ m, 1.5 m and 2 m, $N = 120$ and $\lambda = 0.2$ m. Thus, the element spacings are $0.26\lambda$, $0.39\lambda$ and $0.52\lambda$, respectively for the full-circle array. The boundary of reactive NF and radiative NF is taken to be $(D/\lambda)^{1/3} D/2 \approx 0.2\lambda$ using half-wavelength dipole antenna with antenna size $D = \lambda/2$ [6]. Therefore, the electric field as obtained from (1) is only considered up to $0.2\lambda$ from the array boundary.

### A. Peak Gain

Figure 3 presents the peak value of $|\boldsymbol{E}(x)|$ within the array aperture as the focal point $\boldsymbol{r}_f$ moves along $x$ (i.e., $x_f$) for the three full-circle arrays of different element spacings. The maximum and minimum peak gains appear near the edge (i.e., electric field calculated up to $0.2\lambda$ from the edge) and the center of the circle, respectively. Interestingly, the fluctuations within $2\lambda$ from the center of the array are only around 0.5 dB, indicating stable peak gain despite large coverage areas (inside the array) of up to $\pi(2\lambda)^2 = 4\pi\lambda^2$ m$^2$. Moreover, within $2\lambda$ from the edge of the array, the array gain increases rapidly, by more than 1.5 dB from the minimum value at the array center, for the considered array radii. This is because as the focal point approaches the array edge, the closer elements contribute significantly more to the gain than those farther away. Hence, decreasing the radius also leads to more elements contributing significantly to the gain at any point within the circle, reducing the edge-to-center gain variation.

On the other hand, the half-circle arrays facilitate higher peak gains of up to 5 dB (along $x$) inside the half-circle with the antenna elements (see Fig. 3), relative to their full-circle counterparts. However, this is achieved at the price of up to 12 dB of peak gain loss in the other half-circle. This behavior can be explained by the focal point being at most and at least a distance of $r_c$ away from the nearest element, leading to higher and lower peak gains, respectively. In practice, this result implies that selecting a fixed number of elements closest to the focal point can facilitate significantly higher peak gains, especially for physically larger arrays. In addition, as for the case of the full-circle array, the maximum peak gain occurs at the array's edge, but on the side with the elements for this case.

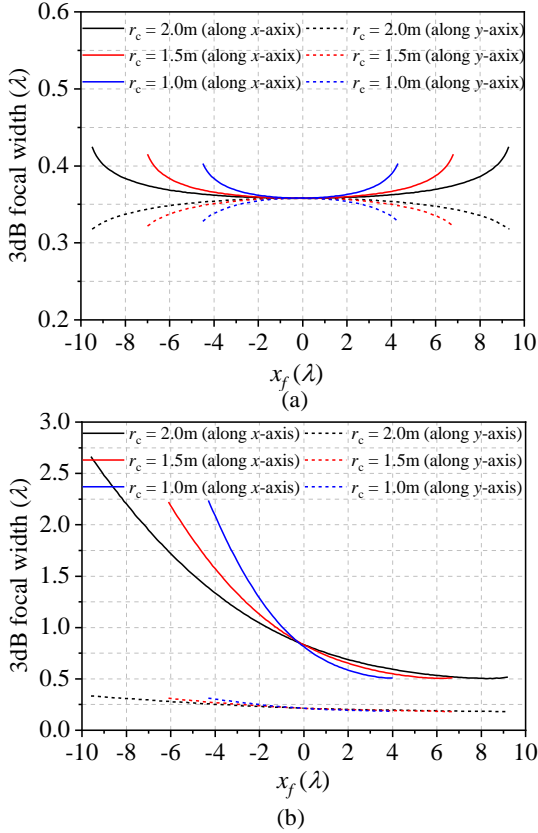

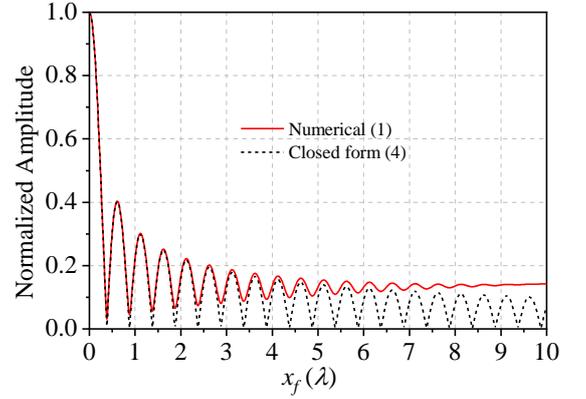

Fig. 5. Numerical (1) and closed form (4) values for normalized electric field magnitude of full-circle array.

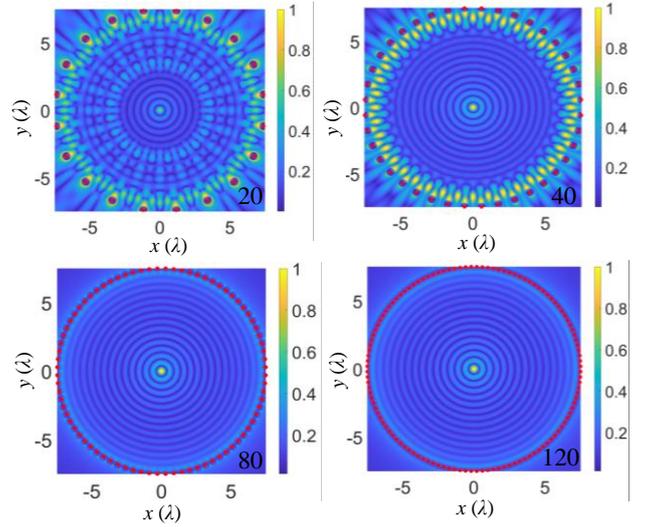

Fig. 6. Normalized electric field distribution for $N$ = 20, 40, 80, 120, with radius $r_c$ = 1.5 m. Element positions are indicated with red dots.

Fig. 4. 3 dB focal width of UCAs: (a) full circle, (b) half-circle.

Moreover, for a sufficiently large $N$, the gain at (0, 0) is always 2.5 dB higher than the gain at ($-r_c$, 0), i.e., on the left edge of the half-circle array that has no element. This can be observed in Fig. 3 and the proof is provided in Appendix A.

### B. Focal Width

To understand how the focal region evolves inside the array, Fig. 4 presents the 3 dB focal width along both $x$ and $y$ axes as the focal point $x_f$ moves along the $x$-axis, for the three circular arrays of each array type. We only consider the range of $x_f$ where the 3 dB width is at least $0.2\lambda$ from the array edge (i.e., boundary of reactive NF and radiative NF). For the full-circle arrays, the width is symmetrical and very stable in both dimensions between the $\pm 2\lambda$ area seen in Fig. 3 for each $r_c$.

Given the circular symmetry of the full-circle arrays, the stable focal width implies a circular 3 dB contour for the focal region. In fact, for the simple case of $x_f = 0$ (focal point at the array center), it is possible to derive an approximate closed form for the focal region and its vicinity. To this end, (1) can be rewritten as follows

$$\boldsymbol{E}(\Delta) = \boldsymbol{a}_z \sum_{n=1}^{N} \frac{e^{-j2\pi\left(\sqrt{r_c^2 + \Delta^2 - 2r_c\Delta\cos\theta_n} - r_c\right)/\lambda}}{\sqrt{r_c^2 + \Delta^2 - 2r_c\Delta\cos\theta_n}} \quad , \tag{2}$$

where $\Delta$ is the distance along the $x$-axis from the array center at (0, 0). As long as $\Delta \ll r_c$, (2) can be simplified by using the Taylor series expansion and the magnitude of the electric field is given by

$$E(\Delta) = \left| \sum_{n=1}^{N} \frac{e^{-\frac{j\pi\Delta^2}{\lambda r_c}} e^{\frac{j2\pi\Delta\cos\theta_n}{\lambda r_c}}}{\sqrt{r_c^2 + \Delta^2 - 2r_c\Delta\cos\theta_n}} \right| . \tag{3}$$

When $N$ is large, the summation in (3) can be approximated by an integral, which yields a known closed form

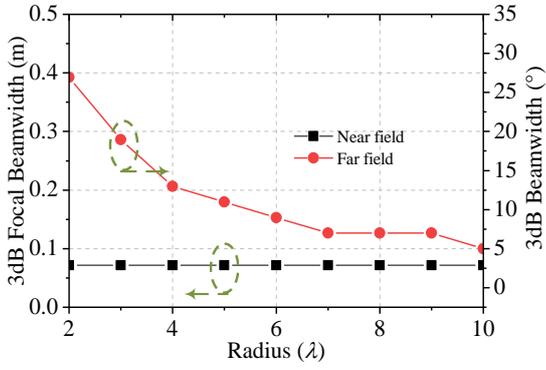

Fig. 7. Comparison of NF focal width and FF beamwidth for different radii of a UCA.

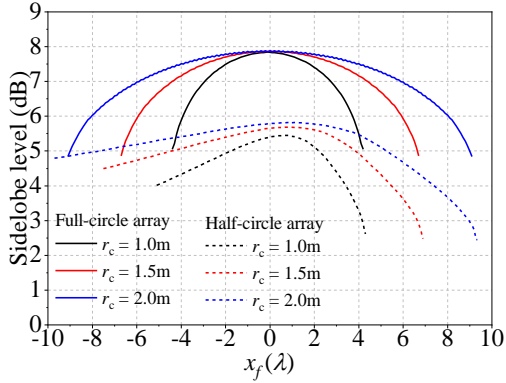

Fig. 8. Sidelobe level of full-circle and half-circle UCAs.

$$E(\Delta) \approx \int_0^{2\pi} \frac{e^{-\frac{j2\pi\Delta\cos\theta}{\lambda r_c}}}{\sqrt{r_c^2 + \Delta^2 - 2r_c\Delta\cos\theta}} d\theta = \frac{\pi}{r_c} J_0\left(\frac{2\pi\Delta}{\lambda}\right), \quad (4)$$

where $J_0(\bullet)$ is the zeroth order Bessel function of the first kind. This means that the focal region follows the shape of a Bessel function, but it is unrelated to the diffraction-free "Bessel beam" in optics [7]. Figure 5 shows the numerical result calculated using (1) and the closed form result in (4). When $\Delta \ll r_c$, the two curves agreed well until about $x = 5\lambda$. The 3 dB focal width obtained from the Bessel function is ~0.36$\lambda$, from the $x$ axis values where the power drops to 3 dB below the peak value. From (4), it can be observed that when $N$ is large enough to apply the Bessel function approximation, the 3 dB focal width is independent of $N$. Figure 6 illustrates the electric field magnitude variation within the full-circle array for different $N$'s, $r_c = 1.5$ m and focal point at the array center, i.e., (0, 0). The electric field in the reactive near-field region for the array, enclosed between the radii of 7.5±0.2$\lambda$, is set to zero since (1) is not valid in this region. As $N$ increases from 20 to 120, the normalized electric field magnitude distribution in and around the main focal region at the center is unchanged, even though the actual magnitude increases due to increasing number of elements. However, away from the center region, the electric field distribution starts to take on the regular pattern seen in Fig. 5 as $N$ increases from 20, primarily due to integral approximation of (5) improving with large $N$. In physical terms, the smaller $N$'s

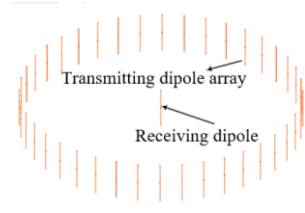

Fig. 9. The simulation model in FEKO for the dipole array ($N = 40$) and receiving dipole in the middle.

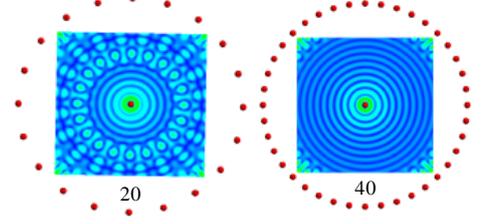

(a)

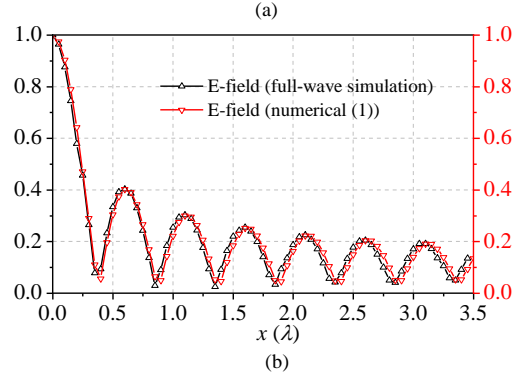

(b)

Fig. 10. (a) Simulated electric field distribution for $N = 20$, 40, with radius $r_c = 1.5$ m. Element positions are indicated with a red point. (b) Comparison of normalized electric field magnitude along $x$-axis for numerical values from (1) and full-wave simulation results with $N = 40$, with radius $r_c = 1.5$ m.

imply larger element spacing (beyond 0.5$\lambda$), resulting in grating lobes.

To further illustrate the difference in the actual array focal (or beam) width between the near and far fields due to changes in the antenna array diameter, Fig. 7 presents the gain beam pattern as the radius of the full circular array increases from 2$\lambda$ to 10$\lambda$, with $N = 120$ at 1.5 GHz. The NF focal width remains constant, whereas the FF beamwidth gradually decreases (hence increasing the array's FF resolution) as the array size increases.

*C. Sidelobes*

The sidelobe levels shown in Fig. 8 only consider sidelobes that are within the array aperture for both full-circle and half-circle arrays. It can be observed that the sidelobe of the full-circle arrays ranges from ~5 dB at the array edge to ~8 dB at the array center, whereas the sidelobe of the half-circle arrays is at best below 6 dB, and the value can drop to as low as 2.5 dB at the edge without array element. Therefore, from the viewpoint of more spatially selective NFF (i.e., low sidelobe in this context), full-circle array is more favorable.

## III. FULL-WAVE VALIDATION

To provide partial validation of the numerical results in the previous section, full-wave simulation was performed for one representative case using the Near field solver in FEKO (version 2022). The solver was chosen to allow for efficient computation given the electrically large arrays.

The case of interest is the full-circle arrays depicted in Fig. 9, with $N = 40$, $\lambda = 0.2$ m and $r_c = 1.5$ m. The focal point was set at $(0, 0)$. For practicality, $z$-polarized half-wavelength dipoles (tuned to 1.5 GHz), modeled by thin cylinders and fed by lumped ports in the middle, were used as the antenna elements. The $N$-element dipole array served as transmitting antennas and a receiving dipole antenna was placed at the center. To extract the electric field magnitude of the receiving dipole (element 0) at the array center (i.e., focal point), only the magnitude of the transmission coefficients $S_{n0}$ ($n = 1,\ldots, N$) of the scattering parameters (S-parameters) is retained, to emulate phase conjugation NFF at the transmitting array. The electric field magnitude is then $|E(x)| \propto \sum_{n=1}^{N} |S_{n0}(x)|$. The electric field distribution with 20/40 elements is shown in Fig.10 (a). The electric field exhibits a standing wave pattern. Although the instantaneous Poynting vector $S = E \times H = 0$ (due to structural symmetry, where $H$ cancels out at the center), the total power is not zero because the electric fields constructively interfere, leading to an enhanced electric field at the center ($1/2|E|^2$). When a very thin half-wave dipole is placed at the center, it does not significantly affect the field distribution. Figure 10 (b) shows that the full-wave simulation results of the normalized electric field magnitude follow closely numerical results (1), up to the second sidelobe. Thereafter, the sidelobes and nulls do not match as well, which can be attributed to antenna scattering not accounted for in the simulation. Moreover, since the array spacing in this case is $\sim 1.18\lambda$, the mutual coupling between the array elements is expected to have a more minor role in causing the observed discrepancy.

Assuming the total input power of the system is normalized to 1, each element receives an equal share of the power, i.e., $1/N$. $N$ is the number of elements. The subsequent analysis focuses on identifying the optimal spacing between elements. This investigation is primarily motivated by the following two considerations:

An increase in the number of elements leads to an improvement in the focusing gain, resulting in higher beamforming performance and enhanced overall system gain.

When the spacing between elements becomes too small, mutual coupling effects become more significant, which inevitably reduces the effective radiated power of each element.

Figure 11 illustrates how the transmission efficiency varies with the number of antenna elements. In this simulation, the antennas are arranged in a circular ring with a radius of 0.5 meters, and the total input power is set to 1 W. The results show that the transmission efficiency reaches its maximum value of 6.688% when the number of elements increases to 21. At this point, the optimal spacing between elements is $0.7452\lambda$. Considering that each dipole experiences a relatively consistent coupling environment, and that the primary coupling effects arise mainly from a few neighboring dipole elements, it is reasonable to use this spacing as a reference. A more accurate approach to determine the optimal spacing between elements—regardless of the ring's radius—is to analyze the system using a coupling matrix [8].

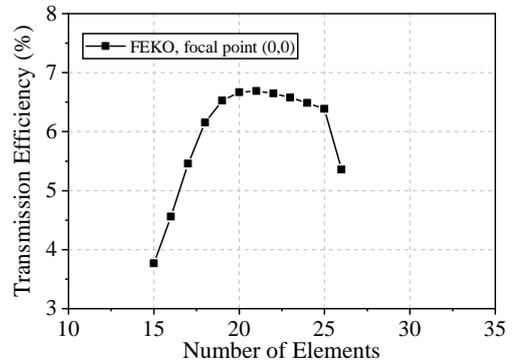

Fig. 11. Transmission efficiency simulated result in FEKO for half-wavelength dipole when the number changes.

## IV. CONCLUSIONS

The NFF properties of simplified LIS setups of circular arrays is investigated in this work for vertically polarized antennas, as the first part of a two-part article. It is found in this part (Part I) that the full-circle array provides stable peak gain and 3 dB focal width when the focal point is within $\pm 2\lambda$ from the array center. On the other hand, the half-circle array shows the potential to apply antenna selection to increase peak gain, but at the cost of larger focal width (stretched in one dimension). In addition, closed form derivations show the electric field distribution closely approximating a Bessel function for sufficiently large number of array elements, a result used to derive the ratio of peak gains between the array center and the edge without elements in the half-circle array. In Part II, a similar study is made for horizontally polarized antennas, where among other conclusions, the array setup with aligned polarization (e.g., all $x$-polarized antennas) was found to yield superior NFF performance. Overall, the results of the two-part article show good potentials of LIS, consisting of array elements that surround the focal point, to provide favorable NFF properties. For future work, it will be interesting to consider more realistic use cases involving scattering objects and multipath propagation and how they impact the expected NFF performance.

## APPENDIX A

Applying phase conjugation and amplitude normalization to (1) gives

$$E(r) = a_z \sum_{n=1}^{N} \frac{1}{|r - r_n|} \quad . \tag{8}$$

For the full-circle array, $r_n = (r_c \cos\theta_n, r_c \sin\theta_n)$. When $N$ is large enough, we can approximate (8) by an integral

$$E(r) = a_z \int_0^{2\pi} \frac{1}{\sqrt{x^2 + r_c^2 - 2xr_c \cos\theta}} d\theta. \quad (9)$$

For the half-circle array with sufficiently large $N$, (9) can be used but for integration region of $(-\pi/2, \pi/2)$. The two field points of interest are and $r_2 = (0,0)$. Calculating the electric fields and taking the ratio

$$E(r_1) = \frac{a_z}{r_c} \ln\left(\frac{\sqrt{2}+1}{\sqrt{2}-1}\right), \quad (10)$$

$$E(r_2) = \frac{a_z \pi}{r_c}, \quad (11)$$

$$|E(r_2)/E(r_1)| = \pi / \ln\left(\frac{\sqrt{2}+1}{\sqrt{2}-1}\right) \approx 2.5 \text{ dB}. \quad (12)$$